\documentclass[prl,twocolumn,amsmath]{revtex4-1}

\usepackage{graphicx}

\usepackage{amssymb,amsfonts,amsmath}

\def \>{\rangle} 
\def \<{\langle} 
 
\def\be{\begin{equation}} 
\def\ee{\end{equation}} 
\def\longrightharpoonup{\relbar\joinrel\rightharpoonup}
\def\longleftharpoondown{\leftharpoondown\joinrel\relbar}

\def\longrightleftharpoons{
  \mathop{
    \vcenter{
      \hbox{
      \ooalign{
        \raise1pt\hbox{$\longrightharpoonup\joinrel$}\crcr
	  \lower1pt\hbox{$\longleftharpoondown\joinrel$}
	  }
      }
    }
  }
}

\newcommand \bea {\begin{eqnarray}} 
\newcommand \eea {\end{eqnarray}}

\begin{document}

\title{ The Energetic Costs of Cellular Computation}

\author{Pankaj Mehta}
\affiliation{Dept. of Physics, Boston University, Boston, MA 02215}
\date{\today}

\author{David J. Schwab }
\affiliation{Dept. of Molecular Biology and Lewis-Sigler Institute, Princeton University, Princeton, NJ 08854}
\date{\today}
\begin{abstract}
Cells often perform computations in order to respond to environmental cues. A simple example is the classic problem, first  considered by Berg and Purcell, of determining the concentration of a chemical ligand in the surrounding media. On general theoretical grounds, it is expected that such computations require cells to consume energy. In particular, Landauer's principle states that energy must be consumed in order to erase the memory of past observations. Here, we explicitly calculate the energetic cost of steady-state computation of ligand concentration for a simple two-component cellular network that implements a noisy version of the Berg-Purcell strategy. We show that learning about external concentrations necessitates the breaking of detailed balance and consumption of energy, with greater learning requiring more energy. Our calculations suggest that the energetic costs of cellular computation may be an important constraint on networks designed to function in resource poor environments, such as the spore germination networks of bacteria.  \end{abstract}

\maketitle

The relationship between information and thermodynamics remains an active area of research despite decades of study \cite{landauer1961irreversibility, berut2012experimental,  bennett1982thermodynamics, del2011thermodynamic}. An important implication of the recent experimental confirmation of Landauer's principle, which relates the erasure of information to thermodynamic irreversibility, is that any irreversible computing device must consume energy \cite{ berut2012experimental,  bennett1982thermodynamics}. The generality of Landauer's argument suggests that this is true regardless of how the computation is implemented. A particularly interesting class of examples relevant to systems biology and biophysics are intracellular biochemical networks that compute information about the external environment. These biochemical networks are ubiquitous in biology, ranging from the quorum-sensing and chemotaxis networks in single-cell organisms to networks that detect hormones and other signaling factors in higher organisms. 

A fundamental question is the relationship between the information processing capabilities of these biochemical networks and their energetic costs \cite{qian2005nonequilibrium,qian2003thermodynamic}. Energetic costs place important constraints on the design of physical computing devices as well as on neural computing architectures in the brain and retina \cite{laughlin2001energy, laughlin1998metabolic, balasubramanian2001metabolically}, suggesting that these constraints may also influence the design of cellular computing networks.

\begin{figure}
\includegraphics[scale=0.4]{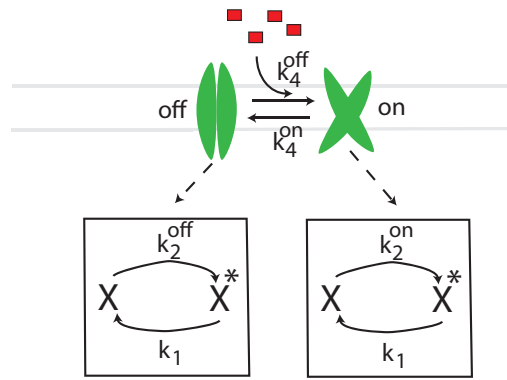}
\caption{{\bf A cellular network for the computation of an external ligand concentration}. External ligands are detected by a receptor that can exist in two conformations: a high-activity  ``on'' state and a low-activity  ``off''. Receptors switch between states at rate $k_{\textrm{off}}$ and  $k_{\textrm{on}}$. Receptors in state $s$=on,off can post-translationally modify (i.e. phosphorylate) a downstream protein at a rate $k_2^s$. This modification is lost (i.e. dephosphorylated) at a rate $k_1$.  }
\label{Fig1}
\end{figure}

The best studied example of a cellular computation is the estimation of a steady-state concentration of chemical ligand in the surrounding environment \cite{berg1977physics, bialek2005physical, endres2009maximum}. This problem was first considered in the seminal paper by  Berg and Purcell who showed that the information a cell learns about its environment is limited by stochastic fluctuations in the occupancy of the receptors that detect the ligand \cite{berg1977physics}. In particular, they considered the case of a cellular receptor that binds ligands at a concentration-dependent rate $k_4^{\textrm{off}}$ and unbinds particles at a rate $k_4^{\textrm{off}}$ (see Fig. \ref{Fig1}) and argued that cells could estimate chemical concentrations by calculating the average time a receptor is bound during a measurement time $T \gg 1$. Recently, however, it was shown that the optimal strategy for a cell is instead to calculate the average duration of the unbound intervals during $T$ or,  equivalently, the total time that the receptor was unbound during $T$. This later computation implements Maximum Likelihood Estimation (MLE) \cite{endres2009maximum}. In these studies, the biochemical networks downstream of the receptors that perform the desired computations were largely ignored because the authors were interested in calculating fundamental limits on how well cells can compute external concentrations. However, calculating energetic costs requires us to explicitly model the downstream biochemical networks that implement these computations \cite{magnasco1997chemical}.

Here, we consider a simple two-component biochemical network that encodes information about ligand concentration in the  steady-state concentration of the activated form of a downstream protein (shown in Figure \ref{Fig1}). Such two-component networks are common in bacteria and often used to sense external signals with receptors phosphorylating a downstream response regulator \cite{laub2007specificity}. Receptors exist in an active `on' state and an inactive `off' state. For simplicity, as in previous works \cite{berg1977physics, bialek2005physical, endres2009maximum}, we assume that the binding affinity in the on state is extremely high so that the ligand-bound receptors are always in the on state and unbound receptors in the off state. Receptors can switch between the off state and on state at a \emph{concentration-dependent} rate $k_4^{\textrm{off}}$ and from the on state to the off state at a concentration-independent rate $k_4^{\textrm{on}}$. Receptors convert a downstream protein from an inactive form $X$ to an active form $X^*$ at a state-dependent rate $k_2^{s}$, with $s=\textrm{on, off}$. The proteins are inactivated at a state-independent rate $k_1$.

The phosphorylation rates in the off state are small but must be nonzero for thermodynamic consistency \cite{beard2008chemical}.  $k_2^{\textrm{off}}$ includes non-specific phosphorylation due to other kinases as well as contributions from the reverse reactions of the phosphotases. The inactivation rate sets the effective measurement time $T \propto k_1^{-1}$ since it is the rate at which information encoded in downstream proteins is lost due to inactivation. In order to compute external concentrations accurately, the measurement time  must be much longer than the typical switching times between receptor states, $k_1 \ll k_4^{\textrm{on}}, k_4^{\textrm{off}}$. We show below that this simple network actually implements a noisy version of the original Berg-Purcell calculation and discuss the relationship between information and power consumption in the context of this network.

\section{Steady-state properties}

The deterministic dynamics of the biochemical network in Figure \ref{Fig1} is captured by simple rate-equations for the mean number of activated proteins. We can augment these equations to account for stochastic fluctuations within the Linear-noise approximation by adding appropriate Langevin noise terms.  In what follows, we assume that proteins are abundant and ignore saturation effects.  The dynamics of the cellular circuit is described by a pair of Langevin equations for the probabilities $p_{\textrm{on}}$ and $p_{\textrm{off}}$  of the receptor to be in the on and off state, respectively, and the number of activated proteins $n$,
\bea
\frac{dp_{\textrm{on}}}{dt} &=& k_4^{\textrm{off}}(1-p_{\textrm{on}})-k_4^{\textrm{on}}p_{\textrm{on}}  + \eta_{\textrm{r}}(t) \\
\frac{d n}{dt} &=& k_2^{\textrm{on}}p_{\textrm{on}} +k_2^{off}(1-p_{\textrm{on}})-k_1n+ \eta_{n}(t).
\label{LangevinEq}
\eea
The variance of the Langevin terms is given by the Poisson noise in each of the reactions  
\bea
\<\eta_n(t)\eta_n(t^\prime)\> &=&( k_2^{\textrm{on}}\bar{p}_{\textrm{on}} +k_2^{off}(1-\bar{p}_{\textrm{on}})+k_1\bar{n})\delta(t-t^\prime)
 \nonumber\\
\<\eta_r(t)\eta_r(t^\prime)\> &=&(k_4^{\textrm{off}}(1-\bar{p}_{\textrm{on}})+k_4^{\textrm{on}}\bar{p}_{\textrm{on}})\delta(t-t^\prime),
\eea
with $\delta(t-t^\prime)$ denoting the Dirac-delta function and barred quantities denoting the mean steady-state values \cite{detwiler2000engineering, mehta2008quantitative}

At steady-state, we can calculate the mean probability and number of proteins by setting the time derivative in Eq. \ref{LangevinEq} equal to zero and ignoring noise terms, yielding
\be
\bar{p}_{\textrm{on}}=1-\bar{p}_{\textrm{off}}=\frac{K_4^{\textrm{off}}}{K_4^{\textrm{off}}+K_4^{\textrm{on}}}
\label{pon}
\ee
and
\be
\bar{n}= (K_2^{\textrm{on}}-K_2^{\textrm{off}}) \bar{p}_{\textrm{on}} +K_2^{\textrm{off}},
\label{nbar}
\ee
where we have defined the dimensionless parameters $K_j^s=k_j^s/k_1$ with $j=2,k$ and $s=\textrm{on,off}$. For the biologically realistic case $K_2^{\textrm{off}} \ll K_2^{\textrm{on}}p_{\textrm{on}}$, as expected, the mean number of proteins is proportional to the kinase activity in the on state times the probability of being in the on state, $\bar{n} \approx K_2^{\textrm{on}}p_{\textrm{on}}$. One can also calculate the variance in protein numbers  (see Appendix) 
\be
\< (\delta n)^2\> = \bar{n}+ (\Delta K_2^{on})^2\frac{\bar{p}_{on} \bar{p}_{off} }{1+K_4^{on}+K_4^{off}}.
\label{varN}
\ee
The first term on the right hand side of the equation arises from Poisson noise in the synthesis and degradation of activated protein, whereas the second term is due to stochastic fluctuations in the state of the receptors. 

In addition to the mean and variance, we will need the full, steady-state probability distributions for $n$  to calculate power consumption. The steady-state distribution can be calculated from the master equation for the probability, $p_s(n)$, of having $n$ active proteins with the receptor in a state $s$,
\begin{multline}
\frac{dp_s(n)}{dt} = k_1(n+1)p_s(n+1) + k_2^sp_s(n-1)\\
+ k_4^{\bar{s}}p_{\bar{s}}(n)- (k_1n+k_2^s+k_4^s)p_s(n)
\label{ME}
\end{multline}
with $\bar{s}= $ off (on)  when $s=$ on (off). At steady-state, the left-hand side of Eq. \ref{ME} is zero and 
\begin{multline}
K_4^{\bar{s}}p_{\bar{s}}(n)=- (n+1)p_{s}(n+1)-K_2^s p_s(n-1)\\
+(n+K_2^s+K_4^s)p_s(n).
\label{SSeq}
\end{multline}
This equation is similar to those found in \cite{iyer2009stochasticity, visco2009statistical} and can be solved via a generating function approach. Define a pair of generating functions,
\be
G_s(n)= \sum_{n=0}^\infty p_s(n) z^n,
\label{defG}
\ee
with $s=\textrm{on, off}$. We can rewrite (\ref{SSeq}) in terms of the generating functions as
\be
\left[ (z-1)\partial_z-K_2^s(z-1)+K_4^s\right]G_s(z)=K_4^{\bar{s}}G_{\bar{s}}(z).
\label{Geq1}
\ee
This equation must be supplemented by initial conditions for the $G_s(z)$. These follow from the observation that $G_{\textrm{on}}(1)=\bar{p}_{\textrm{on}}$ and $G_\textrm{off}(1)=\bar{p}_{\textrm{off}}=1-\bar{p}_{\textrm{on}}$ with $\bar{p}_{\textrm{on}}$ given by Eq. \ref{pon}. As shown in the Appendix, this equations can be solved exactly and yields
\be
G_s(z)=\frac{K_4^{\bar{s}}e^{K_2^s(z-1)}}{K_4^s+K_4^{\bar{s}}} \, _1F_1(K_4^s;1+ K_4^s+K_4^{\bar{s}}; \Delta K_2^s(z-1))
\label{Gsz}
\ee
where $\Delta K_2^s=\Delta K_2^{\bar{s}}- \Delta K_2^s$ and $_1F_1(a;b; z)$ is the confluent hypergeometric function of the first kind. As a check on this expression, we can compute the variance of $n$ directly from Eq. \ref{Gsz} and it is in agreement with (\ref{varN}) (see Appendix).

\begin{figure}
\includegraphics[scale=0.4]{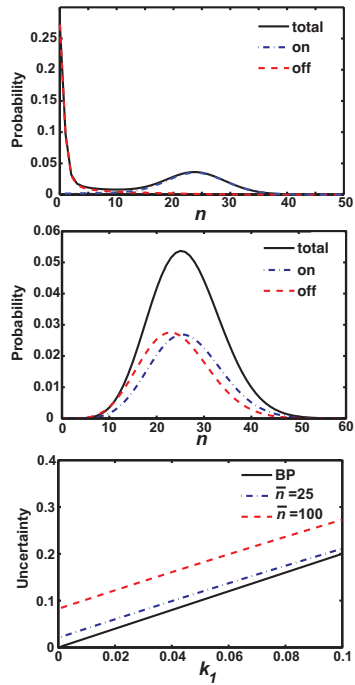}
\caption{ {\bf Top.} Slow switching regime, $k_1 \gg k_4^{\textrm{on}}, k_4^{\textrm{off}}$. Total probability (black solid line),  probability when receptor is in the on state (blue dash-dot line), probability when receptor is in the off state (red dashed line). {\bf Middle.} Fast switching regime, $k_1 \ll k_4^{\textrm{on}}, k_4^{\textrm{off}}$. Total probability (black solid line), probability when receptor is in the on state (blue dash-dot line), probability when receptor is in the off state (red dashed line). {\bf Bottom.} The uncertainty in ligand concentration, $\left(\delta c_{rms}/\bar{c}\right)^2$ as a function of $k_1$ with  mean number of active proteins $\bar{n}=25$  (dashed red line) and $\bar{n}=100$. This can be compared to the Berg-Purcell result (solid black line). Parameters: $k_2^{\textrm{off}}=0.01, k_1^{\textrm{on}}$, $k_4^{\textrm{on}}= k_4^{\textrm{off}}=1$.}   
\label{Figure2}
\end{figure}

Depending on the parameters, the steady-state distributions have two qualitatively distinct behaviors (see Fig. \ref{Figure2}). In the slow switching regime with $k_2^{\textrm{off}} \ll k_2^{\textrm{on}}$ and $k_4^{\textrm{on}}, k_4^{\textrm{off}} \ll k_1 $, receptors switch at rates much  slower than the protein deactivation rate $k_1$.  This gives rise to a bimodal distribution of activated proteins that can be roughly thought of as a superposition of the probability distribution of activated proteins when the receptor is the on and off state.  As $k_2^{\textrm{on}}$ approaches $k_2^{\textrm{off}}$, the distributions in the two states merge and the overall probability distribution becomes unimodal. On the other hand, in the fast switching regime, where $k_4^{\textrm{on}}, k_4^{\textrm{off}} \ll k_1$, the distribution of activated proteins is always unimodal. In this regime, the measurement time $T\propto k_1^{-1}$ is much longer than the average time  a receptor is in the on or off state, and the biochemical network `time-averages' out the stochastic fluctuations in receptor states. In what follows, we restrict our considerations to this latter regime.

\section{Quantifying learning}

The biochemical circuit in Figure \ref{Fig1} ``computes"  the external concentration of  a chemical ligand by implementing a noisy version of Maximum Likelihood Estimation.  As emphasized by Berg and Purcell in their seminal paper \cite{berg1977physics}, the chief obstacle in determining concentration is the stochastic fluctuations in the state of the ligand binding receptors. Berg and Purcell argued that a good measure of how much cells learn is the uncertainty in external concentration as measured by the variance of the estimated concentration $(\delta c)^2$.  Berg and Purcell assumed that the cell computed the average receptor occupancy by time-averaging over a measurement time $T$,  and showed that \cite{berg1977physics}
 \be
 \frac{(\delta c_{\textrm{BP}})^2}{c^2} =\frac{2k_4^{\textrm{on}}}{T\bar{p}_{\textrm{on}}}=2/N_{b},
 \label{deltacBP}
 \ee
with $k_4^{\textrm{off}}=k_+c$, $k_4^{\textrm{on}}=k_-$ and independent of $c$, and $N_b$ the number of binding events during the time $T$.
It was later shown that cells could compute concentration more accurately by implementing Maximum Likelihood Estimation (MLE) with  \cite{endres2009maximum, mora2010limits}
\be
 \frac{(\delta c_{\textrm{ML}})^2}{c^2} =\frac{1}{2} \times \frac{(\delta c_{\textrm{BP}})^2}{c^2} .
\label{deltacMLE}
\ee
The decreased uncertainty is a results from the fact that MLE ignores  noise due to unbinding of ligands from the cell.

To quantify learning in our biochemical circuit, we follow Berg and Purcell and estimate the fluctuations in $(\delta c)^2$ as 
\be
\frac{(\delta c)^2}{c^2} =\left(c \frac{\partial \bar{n}}{\partial c}\right)^{-2} (\delta n)^2,
\label{crms}
\ee
with $(\delta n)^2 = \<n^2\>-\bar{n}^2$. Substituting $k_4^{off}=k_+ c$ and $k_4^{on}=k_-$ and computing the derivative using Eq. \ref{nbar} gives
\be
\left(c \frac{\partial \bar{n}}{\partial c}\right)^2= (\bar{p}_{on}\bar{p}_{off}\Delta K_2)^2.
\label{dercn}
\ee
Substituting Eq. \ref{nbar} and \ref{varN} into Eq. \ref{crms} yields
\be
\frac{(\delta c)^2}{c^2} = \frac{\bar{n}}{(\bar{p}_{on}\bar{p}_{off}\Delta K_2)^2}+ \frac{1}{(\bar{p}_{on} \bar{p}_{off}) (1+K_4^{on}+K_4^{off})}.
\label{crmsfinal}
\ee
Similar to the linear noise calculation, the first term on the right-hand side arises from the Poisson fluctuations in activated protein number whereas the second term comes from the stochastic fluctuations in the state of receptors. Figure \ref{Figure2} shows the uncertainty, ${(\delta c)^2}/{c^2}$, as a function of the degradation rate of activated protein, $k_1$ when $\bar{n}=25$ and $\bar{n}=100$ and $k_2^{\textrm{on}} \gg k_2^{\textrm{off}}$. 

By identifying the degradation rate with the inverse measurement time,  $k_1=2T^{-1}$, we can also compare the results with Berg Purcell. The factor of two is due to the slight difference in how the variance of  the average receptor occupancy is calculated for a biochemical network when compared to the original Berg Purcell calculation \cite{mora2010limits}. As shown in Fig. \ref{Figure2}, when $\bar{n}$ is increased, the Poisson noise in protein production is suppressed and the performance of the cellular network approaches that due to Berg-Purcell. To make the connection with Berg-Purcell more explicit, it is helpful to rewrite Eq. \ref{crmsfinal} in terms of the average number of binding events, $N_b$, during $T$ 
\be
\frac{(\delta c)^2}{c^2} =\frac{ \bar{n}}{(\bar{n}-K_2^{off})^2p_{off}^2}+ \frac{2}{N_{bind}}\left( 1-\frac{k_1}{k_4^{on}+k_4^{off}+k_1}\right).
\ee
When the measurement time is much longer than the timescale for fluctuations in the receptor number, $k_4^{\textrm{on,off}} \gg k_1$ (equivalently $K_2^{\textrm{on}} \gg K_2^{\textrm{off}} \gg 1$), and the average number of activated proteins is large, $\bar{n} \gg K_2^{off} \gg 1$, the expression above reduces to  ${(\delta c)^2}/{c^2} \approx 2/N_{b}$ in agreement with the Eq. \ref{deltacBP}. 

\section{Power consumption and entropy production}

\begin{figure}
\includegraphics[scale=0.4]{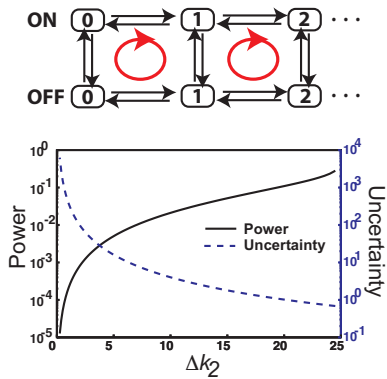}
\caption{  {\bf Top.} The probabilistic Markov process underlying the circuit in Fig. \ref{Fig1}.  Any non-zero cyclic flux (depicted in red) results in entropy production and power consumption. {\bf Bottom.} Power consumption (solid black line)  and uncertainty (dashed purple line) as a function of $\Delta k_2= k_2^\textrm{on}-k_2^\textrm{off}$ when $\bar{n}=25$, $k_4^{\textrm{on}}= k_4^{\textrm{off}}=1$. }   
\label{Figure3}
\end{figure}

We now compute the energy consumed by the circuit in Figure \ref{Fig1} as a function of the kinetic parameters. To do so, we exploit the fact that dynamics of the circuit can be thought of as a nonequilbrium Markov process (see Fig. \ref{Figure3}). A nonequilibrium steady-state (NESS) necessarily implies the breaking of detailed balance in the underlying Markovian dynamics and therefore a non-zero entropy production rate.  The entropy production rate is precisely the amount of power consumed by the biochemical circuit to maintain the nonequilibrium steady-state. Thus, by calculating the entropy production rate as function of kinetic parameters, we can calculate the power consumed by the biochemical network implementing the computation.

Consider a general Markov process with states labeled by $\sigma$ and transition probability from $\sigma$ to $\sigma^\prime$ given by $k(\sigma, \sigma^\prime)$. Defining the steady-state probability of being in state $\sigma$ by $P_{\sigma}$, the entropy production rate for a NESS is given by  \cite{lebowitz1999gallavotti},
\be
\frac{dS}{dt}= \sum_{\sigma, \sigma^\prime}P(\sigma) k(\sigma, \sigma^\prime) \log{\frac{k(\sigma, \sigma^\prime)}{k(\sigma^\prime, \sigma)}},
\label{defA}
\ee
For the biochemical network described by Eq. \ref{ME}, this becomes,
\begin{widetext}
\be
\frac{dS}{dt} = \sum_{s=\textrm{on,off}, n} p_s(n) \left[ k_2^s\log{\frac{k_2^s}{k_1(n+1)}} 
+ k_1n\log{\frac{k_1n}{k_2^s}}+k_4^s\log{\frac{k_4^s}{k_4^{\bar{s}} }} \right]
\label{EP1}
\ee
\end{widetext}
Since the receptors are in thermodynamic equilibrium, from detailed balance we know that
\be
\sum_{s,n} p_s(n) k_4^s \log{\frac{k_4^s}{k_4^{\bar{s}}}}= 0,
\ee
so that 
\be
\frac{dS}{dt} = k_1 \sum_{s=on,off}\sum_n p_s(n) \left( K_2^s\log{\frac{K_2^s}{n+1}}- n \log{\frac{K_2^s}{n}} \right),
\label{AvgA}
\ee
where $K_2^s= k_2^s/k_1$. The steady-state distributions $p_s(n)$ follow from Eq. \ref{Gsz}. The physical content of this expression is summarized in Fig. \ref{Figure3}. The expression states that any non-zero cyclic flux must necessarily produce entropy. If it didn't, one would have a chemical version of a perpetual motion machine. Figures \ref{Figure3} and \ref{Figure4} show the power consumption as a function of $\Delta K_2=K_2^{\textrm{on}}-K_2^{\textrm{off}}$ and $k_1$. Notice that the power consumption tends to zero as both these parameters go to zero. We cannot, however, set $k_1=0$ identically because there then no longer exists a steady-state distribution.

\section{Energetics, Information, and Landauer's Principle}

 \begin{figure}[h]
\includegraphics[scale=0.4]{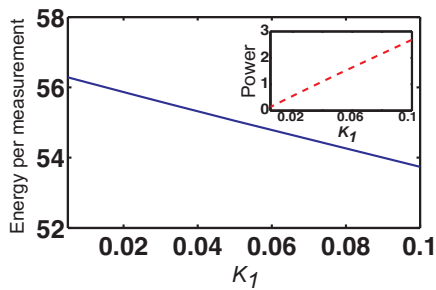}
\caption{Total energy per independent measurement (Power $\times k_1^{-1}$) as a function of $k_1$ when $\bar{n}=25$. (Inset) Power as a function of $k_1$.}   
\label{Figure4}
\end{figure}

We now highlight the fundamental connection between the energy consumed by the network and the information the network learns about the environment, and briefly discuss its relation to Landauer's principle. First, note that learning information about the environment requires energy consumption by the network. This can be seen in Fig. \ref{Figure3} which shows that as $\Delta k_2 \rightarrow 0$, the uncertainty about the concentration tends to infinity. This can be made concrete by noting that the entropy production (Eq. \ref{AvgA}) is zero if and only if $\Delta k_2=0$ (see Appendix).  In conjunction with Eq. \ref{crmsfinal}, which diverges as $\Delta k_2 \rightarrow 0$, this implies that learning requires consuming energy. Physically, in the limit where $\Delta k_2=0$, the dynamics of the Markov process in Fig. \ref{Figure3} become ``one-dimensional" and the dynamics obey detailed balance. At the same time, in this limit, the number of downstream proteins becomes insensitive to external ligand concentrations, since all information about concentration is contained in the relative probabilities of being in the on or off state.

Second, as shown in Fig. \ref{Figure4}, the power consumption of the circuit tends to zero as $k_1 \rightarrow 0$. This is consistent with Landauer's principle: entropy production stems from erasing memory in a computing device.  The number of activated proteins serves as a memory of ligand concentration which is erased at a rate $k_1$. Thus, as the erasure rate of the memory tends to zero, the device consumes less energy, as expected. Yet despite the fact that the power consumption tends to zero as $k_1$ decreases, the total energy consumed per measurement, namely the power times the measurement time, $T\simeq 2k_1^{-1}$, still increases (see Fig. \ref{Figure4}). Thus, learning more requires consuming more total energy despite the fact that power consumption is decreasing. In effect, one is approaching the reversible computing limit where memory is erased very infrequently. Note, however, that when erasure is performed infinitely slowly, $k_1=0$, the system no longer has a NESS and our formalism does not apply.

Finally, we note that one of the important open problems in our understanding are the constraints placed on the measurement time $T$. In principle, cells can always learn more by measuring the environment for longer periods of time. However, in practice, these measurement times tend to be quite short. There are a number of constraints that can set this measurement time including rotational diffusion \cite{berg1977physics} and the restrictions placed on motility. Here, we highlight another restriction that may be important in resource-starved environments: sensing external concentration necessarily requires cells to consume energy. 

\section{Discussion and Conclusion}

Cells often perform computations using elaborate biochemical networks that respond to environmental cues. One of the most common simple networks found in bacteria are two-component networks where a receptor phosphorylates a downstream response regulator \cite{laub2007specificity}. In this work, we have shown that these simple two-component networks can implement a noisy version of the Berg-Purcell strategy to compute the concentration of external ligands. Furthermore, by mapping the dynamics of the biochemical network to Nonequilibrium Steady-States in Markov processes, we explicitly derived expressions for the power consumed by the network and showed that learning requires energy consumption. Taken together, these calculations suggest that, much like man-made and neural computing \cite{laughlin2001energy, laughlin1998metabolic, balasubramanian2001metabolically, bennett1982thermodynamics}, energetic considerations may place important constraints on the design of biochemical networks that implement cellular computations. They also suggest a fundamental relationship between the efficiency of cellular computing and the energy consumption.

Bacterial cells such as {\it Bacillus subtilis} can sporulate during times of environmental stress and remain metabolically dormant for many years. While sporulation is relatively well understood, the reverse process of germination is much more difficult to study. One current model for how a spore knows when to germinate in response to external cues involves integrating the signal and triggering commitment when an accumulation threshold is reached \cite{yi2011synergism, indest2009workshop}. This corresponds to the limit of vanishingly small $k_1$ in our model, so that power consumption is minimized at the expense of retaining the entire integrated signal. Our results indicate that this behavior may be due to the extreme energetic constraints imposed on a metabolically dormant spore, rather than an evolutionarily optimized strategy.

An important insight of this work is that even a simple Berg-Purcell strategy for sensing external concentrations requires the consumption of energy. It is likely that more complicated strategies that increase how much cells learn, such as Maximum Likelihood, require additional energetic inputs. For example, it was argued in \cite{mora2010limits} that MLE can be implemented by a network similar to the perfect adaptation network where bursts are produced in response to binding events. These bursts break detailed balance and therefore require energy consumption. It will be interesting to investigate further how the trade-off between learning and energy consumption manifests itself in the design
 of computational strategies employed by cells.

In this work, we restricted ourselves to the simple case where cells calculate the steady-state concentration of an external signal. In the future, it will be useful to generalize this to other computations such as responding to temporal ramps \cite{mora2010limits} and spatial gradients \cite{endres2008accuracy, hu2010physical}. It will also be interesting to understand how to generalize the considerations here to arbitrary biochemical networks. An important restriction on our work is  that we reduced our considerations to nonequilibrium steady-states. It will be interesting to ask how to generalize the work here to biochemical networks with a strong temporal component. 

\section{Acknowledgements}
PM and DJS would like to thank the Aspen Center for Physics, where this work was initiated.  We are especially grateful to Thierry Mora for clarifying the relationship between the rate $k_1$ and the average integration time $T$. This work was partially supported by NIH Grants K25GM086909 (to P.M.). DS was partially supported by DARPA grant HR0011-05-1-0057 and NSF grant PHY-0957573.

\bibliography{refsmain}   
\onecolumngrid

\appendix

\section{Appendix: Variance from the Linear-noise Approximation}

 Consider the Langevin equations for the ODE's
\bea
\frac{dp_{\textrm{on}}}{dt} &=& k_4^{\textrm{off}}(1-p_{\textrm{on}})-k_4^{on}p_{\textrm{on}} + \eta_{\textrm{r}}(t)  \\
\frac{dn}{dt} &=& k_2^{on}p_{\textrm{on}} +k_2^{\textrm{off}}(1-p_{\textrm{on}})-k_1n +\eta_n(t)
\eea
We can linearize (with bar denoting average) to get
\bea
\frac{d \delta p_{\textrm{on}}}{dt} &=&  -(k_4^{\textrm{off}}+k_4^{on}) \delta p_{\textrm{on}} - \tau_P^{-1} \delta p_{\textrm{on}} +\eta_{\textrm{r}}(t) \\
\frac{d \delta n}{dt} &=& (k_2^{on}-k_2^{\textrm{off}}) \delta p_{\textrm{on}}-k_1 \delta n +\eta_n
\eea
where  
\be
\< \eta_P (t) \eta_P (t) \> = k_4^{\textrm{off}} (1-\bar{p}_{\textrm{on}})+k_4^{on}\bar{p}_on =2 (k_4^{\textrm{off}}+k_4^{on})\bar{p}_{\textrm{on}}(1-\bar{p}_{\textrm{on}}) =2 \tau_P^{-1}\bar{p}_{\textrm{on}}(1-\bar{p}_{\textrm{on}})
\ee
and $\<\eta_n (t) \eta_n(t)\>=2 k_1 \bar{n}$.
Now we Fourier transform the above equations and use the fact that 
\be
\<(\delta n)^2 \>=\int \frac{d\omega}{2\pi} \<\delta \hat{n}(\omega) \delta \hat{n}^*(\omega)\>
\ee
to get
\be
\<(\delta n)^2 \> = \bar{n} + \frac{(\Delta k_2^{on})^2}{k_1^2} \frac{k_1}{k_1+\tau_P^{-1}}=\bar{n}+ (\Delta K_2^{on})^2\frac{\bar{p}_{\textrm{on}} \bar{p}_{\textrm{off}} }{1+K_4^{on}+K_4^{\textrm{off}}},
\ee
where we have used $K_i^s=k_i^s/k_1$. 

\section{Appendix: Generating function for probability distribution}

We start with the equation for the  generating functions in the main text,
\be
\left[ (z-1)\partial_z-K_2^s(z-1)+K_4^s\right]G_s(z)=K_4^{\bar{s}}G_{\bar{s}}(z).
\label{Geq1}
\ee
Adding the equations for $s=$on, off  and dividing through by $(z-1)$ gives
\be
(\partial_z -K_2^{\textrm{on}})G_{\textrm{on}}(z)= -(\partial_z -K_2^{\textrm{off}})G_{\textrm{off}}(z).
\label{Geq2}
\ee

To proceed further, it is useful to define the quantity $H_s(z)$ related to the generating functions
$G_s(z)$ by
\be
G_s(z)= e^{K_2^s z}H_s(z).
\label{defH}
\ee
It is clear that 
\be
(\partial_z -K_2^{s})G_{s}(z)= e^{K_2^s z}\partial_z H_s(z)
\ee
Thus, we can rewrite (\ref{Geq2}) as
\be
e^{K_2^{on}z}\partial_z H_{\textrm{on}}(z)= -e^{K_2^{\textrm{off}}z}\partial_z H_{\textrm{off}}(z).
\label{Heq1}
\ee
Similarly, we can rewrite (\ref{Geq1}) as 
\be
(z-1)e^{K_2^s z} \partial_z H_s(z) + K_4^s e^{K_2^s z}H_s(z)= K_4^{\bar{s}} e^{K_2^{\bar{s}} z} H_{\bar{s}}(z)
\ee
Multiplying the equation by $e^{-K_2^{\bar{s}} z}$, taking the derivative withe respect to $z$, substituting (\ref{Heq1}),  and defining $\Delta K_2^s =K_2^{\bar{s}}-K_2^ss$ one has
\be
  \partial_z H_s(z) - \Delta K_2^s (z-1)  \partial_z H_s(z) + (z-1)\partial_z^2H_s(z)- K_4^s\Delta K_2^s  H_s(z) + K_4^s \partial_z H_s(z)= -K_4^{\bar{s}} \partial_z H_s(z)
\ee
Regrouping terms one has
\be
(z-1)\partial_z^2 H_s(z)+ (1-\Delta K_2^s (z-1) + K_4^s+ K_4^{\bar{s}}) \partial_z H_s(z) - K_4^s \Delta K_2^s H_s(z)=0.
\label{Heq2}
\ee
Defining
\be
u= \Delta K_2^s (z-1),
\ee
one can rewrite (\ref{Heq2}) in terms of $u$ as
\be
u\partial_u^2 H_s(u)+(1+K_4^{s}+K_4^{\bar{s}}-u)\partial_u H_s(z)- K_4^s H_s(u)=0.
\label{Heq3}
\ee
This is just the confluent hypergeometric equation. We can immediately write the solutions in terms of confluent hypergeometric functions of the first and second kind. In particular, the general solution to this equation is given by (assuming that we need a power series in u with integer cofficients) the confluent geometric function:
\be
H_s(u)= c_s\, {}_1 F_1 (K_4^s; 1+K_4^s+K_4^{\bar{s}}; u),
\ee
with $c_s$ a constant of integration. Thus, we have
\be
G_s(z)= c_s e^{K_2^s z} {}_1 F_1 (K_4^s; 1+K_4^s+K_4^{\bar{s}}; \Delta K_2^s (z-1)).
\label{finalG}
\ee
To determine the constants, notice that 
\be
G_s(1)= c_se^{K_2^s}=\bar{p}_s=\frac{K_2^{\bar{s}}}{K_2^s+K_2^{\bar{s}}}
\ee
so that 
\be
 c_s= \frac{e^{-K_2^s} K_2^{\bar{s}}}{K_2^s+K_2^{\bar{s}}}.
\ee
This gives the final expression in the main  text
\be
G_s(z)=\frac{K_4^{\bar{s}}e^{K_2^s(z-1)}}{K_4^s+K_4^{\bar{s}}} \, _1F_1(K_4^s;1+ K_4^s+K_4^{\bar{s}}; \Delta K_2^s(z-1)).
\ee

\section{Appendix: Variance from full probability distribution}
We can calculate the variance $(\delta n)^2$ directly from the generating functional. This serves as a check on our formalism and our expressions. In terms of the generating functions,
\be
(\delta n)^2= \sum_s (\partial_z z \partial_z G_s(z))|_{z=1} -\bar{n}^2= \sum_s ( z \partial_z^2 G_s(z))|_{z=1}+ \bar{n}-\bar{n}^2
\ee
where we have used 
\be
z\partial_z \,_1F_1(a,b ;z) =z \frac{b}{a}\,_1F_1(a+,b+ ;z).
\ee
Taking the second derivative and setting $z=1$ yields
\begin{align}
 ( z \partial_z^2 G_s(z))|_{z=1}= \frac{K_4^s}{K_4^s+K_4^{\bar{s}}} \left[ (K_2^s)^2+ \frac{ \Delta K_2^s K_4^s}{1+K_4^s+K_4^{\bar{s}}}\left( 2K_2^s + \frac{\Delta K_2^s}{2+K_4^s+K_4^{\bar{s}}} \right)\right].
\end{align}
After some algebra, one has
\be
 \sum_s ( z \partial_z^2 G_s(z))|_{z=1}= \bar{p}_{\textrm{on}} (K_2^{on})^2+ \bar{p}_{\textrm{off}} (K_2^{\textrm{off}})^2 - \bar{p}_{\textrm{on}}\bar{p}_{\textrm{off}} (\Delta K_2^s)^2 \frac{K_4^{on}+K_4^{\textrm{off}}}{1+K_4^{on}+K_4^{\textrm{off}}}.
\ee
This gives
\begin{align}
(\delta n)^2 &= \bar{n}+  \bar{p}_{\textrm{on}} (K_2^{on})^2+ \bar{p}_{\textrm{off}} (K_2^{\textrm{off}})^2 - \bar{p}_{\textrm{on}}\bar{p}_{\textrm{off}} (\Delta K_2^s)^2 \frac{K_4^{on}+K_4^{\textrm{off}}}{1+K_4^{on}+K_4^{\textrm{off}}} - \bar{n}^2 \\
&=\bar{n}+  \bar{p}_{\textrm{on}} (K_2^{on})^2+ p_{\textrm{off}} (K_2^{\textrm{off}})^2 - \bar{p}_{\textrm{on}}\bar{p}_{\textrm{off}} (\Delta K_2^s)^2 \frac{K_4^{on}+K_4^{\textrm{off}}}{1+K_4^{on}+K_4^{\textrm{off}}} - (\bar{p}_{\textrm{on}} K_2^{on}+\bar{p}_{\textrm{off}}K_2^{\textrm{off}})^2\\
&= \bar{n}+ \bar{p}_{\textrm{on}}\bar{p}_{\textrm{off}}\frac{( K_2^{on}-K_2^{\textrm{off}})^2}{1+K_4^{on}+K_4^{\textrm{off}}}.
\end{align}

\subsection{Appendix: Energy consumption is required for signaling}
We show is that $\frac{dS}{dt}=0$ if and only if $\Delta K_2^s=0$. To do so we will use the expressions for the probability distribution (\ref{Gsz}) and average entropy production (\ref{AvgA}). Let us start by writing (\ref{AvgA}) as
\begin{align}
k_1^{-1} \frac{dS}{dt}= - \left( \<n\>_{\textrm{on}} \log K_2^{on}+\<n\>_{\textrm{off}} \log K_2^{\textrm{off}} \right)+\sum_n \left( K_2^{on} p_{\textrm{on}}(n) \log K_2^{on} +   K_2^{\textrm{off}} p_{\textrm{off}}(n) \log K_2^{\textrm{off}} \right) \\
+ \sum_n \left( [p_{\textrm{on}}(n+1)-K_2^{on}p_{\textrm{on}}(n)]\log{(n+1)} \right) +\sum_n \left( [p_{\textrm{off}}(n+1)-K_2^{\textrm{off}} p_{\textrm{off}}(n)]\log{(n+1)} \right),
\label{ADB}
\end{align}
where $\<n\>_s= \sum_n n p_s(n)$ for $s=\textrm{on, off}$.
We will now examine each of these terms one by one. Notice that 
\be
\<n\>_s= \partial_z G_s(z=1)= \frac{K_2^{\bar{s}} K_4^{\bar{s}}}{K_4^s+ K_4^{\bar{s}}},
\ee
so we have that
\be
 \left( \<n\>_{\textrm{on}} \log K_2^{on}+\<n\>_{\textrm{off}} \log K_2^{\textrm{off}} \right)= \sum_s \frac{K_2^{\bar{s}} K_4^{\bar{s}}}{K_4^s+ K_4^{\bar{s}}}\log{K_2^{s}}.
\ee
Notice that since $\sum p_s(n)= G_s(1)$, the second term in the sum above is just
\be
\sum_n \left( K_2^{on} p_{\textrm{on}}(n) \log K_2^{on} +   K_2^{\textrm{off}} p_{\textrm{off}}(n) \log K_2^{\textrm{off}} \right) =  \sum_s \frac{K_2^{s} K_4^{\bar{s}}}{K_4^s+ K_4^{\bar{s}}}\log{K_2^{s}}.
\ee
Substituting these expressions into (\ref{ADB}) yields
\begin{align}
\frac{A_p}{k_1} &= -\sum_s \frac{K_4^s \Delta K_2^s}{K_4^{s}+K_4^{\bar{s}}} \log K_2^s 
+ \sum_n \left( [p_{\textrm{on}}(n+1)-K_2^{on}p_{\textrm{on}}(n)]\log{(n+1)} \right)\\
& +\sum_n \left( [p_{\textrm{off}}(n+1)-K_2^{\textrm{off}} p_{\textrm{off}}(n)]\log{(n+1)} \right).
\label{ADB}
\end{align}
In order for this to equal zero, we know that 
\be
p_s(n+1)=\frac{K_2^s}{n+1} p_s(n),
\ee
since each of the terms in the sum above must equal zero individually. This implies that the generating functional in detailed balance, $G_s^{DB}(z)$, must be given by
\be
G_s^{DB} \propto e^{K_2^s z}.
\ee
Comparing with (\ref{Gsz}) implies that $\Delta K_2^s=0$. In addition, the first term of (\ref{ADB}) also disappears in this case. Thus, we have detailed balance if and only if 
$\Delta K_2^s =0$. 
\end{document}